\documentclass[prl,twocolumn]{revtex4}

\usepackage[utf8]{inputenc}
\usepackage[english]{babel}
\usepackage[T1]{fontenc}
\usepackage{amsfonts}
\usepackage[dvips]{graphicx}
\usepackage[usenames]{color}
\usepackage{amssymb,amsmath}

\DeclareMathOperator{\Tr}{Tr}

\begin{document}
\title{Optimal Quantum Tomography of States, Measurements, and Transformations}

\author{A. Bisio} 
\affiliation{Quit group,
  Dipartimento di Fisica ``A. Volta'', via Bassi 6, I-27100 Pavia,
  Italy and CNISM.} 

\author{G. Chiribella} 
\affiliation{Quit group,
  Dipartimento di Fisica ``A. Volta'', via Bassi 6, I-27100 Pavia,
  Italy and CNISM.} 

\author{G. M. D'Ariano} 
\affiliation{Quit group,
  Dipartimento di Fisica ``A. Volta'', via Bassi 6, I-27100 Pavia,
  Italy and CNISM.} 

\author{S. Facchini} 
\affiliation{Quit group,
  Dipartimento di Fisica ``A. Volta'', via Bassi 6, I-27100 Pavia,
  Italy and CNISM.} 

\author{P. Perinotti} 
\affiliation{Quit group,
  Dipartimento di Fisica ``A. Volta'', via Bassi 6, I-27100 Pavia,
  Italy and CNISM.} 

\begin{abstract}
  We present the first complete optimization of quantum tomography, for states, POVMs, and various
  classes of transformations, for arbitrary prior ensemble and arbitrary representation, giving
  corresponding feasible experimental schemes.
\end{abstract}

\maketitle

\newcommand{\ket}[1]{| #1 \rangle}
\newcommand{\bra}[1]{\langle #1 |}
\newcommand{\Ket}[1]{| #1 \rangle \! \rangle}
\newcommand{\Bra}[1]{\langle \! \langle #1 |}
\newcommand{\BraKet}[2]{\langle \! \langle #1 | #2 \rangle  \! \rangle}
\newcommand{\KetBra}[2]{\Ket{#1} \Bra{#2}}
\newcommand{\braket}[2]{\langle #1 | #2 \rangle}
\newcommand{\ketbra}[2]{\ket{#1} \bra{#2}}
\newcommand{\hilb}[1]{\mathcal{#1}}

\newcommand{\sumint}{\int\hspace{-16pt}\sum}

\def\N#1{\Vert{#1}\Vert}
\def\refp#1{(\ref{#1})}
\def\>{\rangle} \def\<{\langle}

A crucial issue in quantum information theory is the precise
determination of states and processes. The procedure by which this
task can be accomplished is known as \emph{quantum tomography}
\cite{libroparis,tomoperation,tomaccone}.

The most general quantum measurement is described by a POVM, namely a collection of positive
operators $P_i \in \mathcal{B}(\hilb{H})$ satisfying the normalization $\sum_i P_i = I$
\cite{holevo}.  The probability distribution of the outcome $i$ of the measurement is provided by
the Born statistical formula
\begin{equation}
 p_i = \Tr[\rho P_i].
\end{equation}
Tomographing an unknown state $\rho$ of a quantum system means performing a suitable POVM $\{ P_i
\}$ such that every expectation value can be evaluated from the probability distribution $p_i =
\Tr[\rho P_i]$.  In particular the expectation value of an operator $A$ can be obtained when it is
possible to expand $A$ over the POVM as follows
\begin{equation}\label{exp-obs}
A = \sum_i f_i[A] P_i,
\end{equation}
$f_i[A]$ denoting suitable expansion coefficients. The expectation of
$A$ is then obtained as $\< A \> = \sum_i f_i[A] \< P_i \>$. When
expansion (\ref{exp-obs}) holds for all operators ${\mathcal
  B}(\hilb{H})$---i.~e.  $\mathcal{B}(\hilb{H})= {\sf span} \{ P_i \}$---the
POVM is called \emph{informationally complete}
\cite{prugovecki,busch}.
 
It is convenient to associate every operator $A \in
\mathcal{B}(\hilb{H})$ to a bipartite vector in
$\hilb{H}\otimes\hilb{H}$ in the following way
\begin{equation}\label{eq-corr}
A=\sum_{m,n=1}^dA_{mn}\ketbra{m}{n} \leftrightarrow \Ket{A}=\sum_{m,n=1}^dA_{mn}\ket{m}\ket{n}.
\end{equation}
Information-completeness of the POVM along with convergence of the
series (\ref{exp-obs}) rewrite as follows
\begin{equation}\label{eq-frame}
a\N{A}^2_2\leq\sum_{i=1}^N|\BraKet{P_i}{A}|^2\leq b\N{A}_2^2,\quad A\in{\mathcal \mathcal{B}(\hilb{H})},
\end{equation}
with $0 < a \leq b < \infty$. Sets of vectors $ \Ket{P_i} $ satisfying condition
(\ref{eq-frame}) are known as \emph{frames}
\cite{frame1}. This condition is equivalent to invertibility
of the \emph{frame operator} $ F = \sum_i \KetBra{P_i}{P_i}$.  The
expansion in Eq. (\ref{exp-obs}) can be written as follows
\begin{equation}
\Ket{A} = \sum_i\BraKet{D_i}{A} \Ket{P_i}, 
\end{equation}
in terms of a \emph{dual frame} $\{D_i\}$, namely a set of
operators  satisfying the identity $ \sum_i
\KetBra{P_i}{D_i}=I$.  For linearly dependent frame $\{P_i\}$ the dual
$\{D_i\}$ is not unique.

The request for the POVM $\{ P_i \}$ to be informationally complete
can be relaxed if we have some prior information about the state
$\rho$. If we know that the state belongs to a given subspace
$\mathcal{V} \subseteq \mathcal{B}(\hilb{H})$ the expectation value is
\begin{equation}\label{eq-obsproj}
\< A \> = \BraKet{\rho}{A} = \Bra{\rho}Q_\mathcal{V}\Ket{A}
\end{equation}
$Q_\mathcal{V}$ orthogonal projector on $\mathcal{V}$, whence the set
$\{ P_i \}$ is required to span only $\mathcal{V}$.

For the estimation of the expectation $\< A \>$ of an observable $A$, optimality means minimization
of the cost function given by the variance $\delta(A)$ of the random variable $\BraKet{D_i}{A}$ with
probability distribution $\Tr[\rho P_i]$, namely
\begin{equation}
\delta(A) := \sum_i |\BraKet{D_i}{A}|^2 \Tr[\rho P_i] - |\Tr[\rho A]|^2.
\end{equation}
In a Bayesian scheme the state $\rho$ is randomly drawn from an
ensemble $\mathcal{S} = \{\rho_k, p_k \}$ of states $\rho_k$ with
prior probability $p_k$, with the variance averaged over
$\mathcal{S}$, leading to
\begin{equation}
 \delta_\mathcal{S}(A) := \sum_i |\BraKet{D_i}{A}|^2 \Tr[\rho_\mathcal{S} P_i] - \sum_k p_k |\Tr[\rho_k A]|^2
\end{equation}
where $\rho_\mathcal{S}=\sum_k p_k \rho_k$. Moreover, a priori we can
be interested in some observables more than other ones, and this can
be specified in terms of a weighted set of observables $\mathcal{G} =
\{A_n , q_n \}$, with weight $q_n>0$ for the observable $A_n$. Averaging
over $\mathcal{G}$ we have
\begin{equation}\label{eq-figmersg}
\delta_{\mathcal{S},\mathcal{G}} := \sum_i \Bra{D_i}G\Ket{D_i} \Tr[\rho_\mathcal{S} P_i] - \sum_{k,n} p_k q_n |\Tr[\rho_k A_n]|^2
\end{equation}
where $G=\sum_n q_n \KetBra{A_n}{A_n}$. The weighted set $\mathcal{G}$ yields a representation of
the state, given in terms of the expectation values. The representation is faithful when $\{A_n\}$
is an operator frame, e.~g. when it is made of the dyads $|i\>\< j|$ corresponding to the matrix elements
$\<j|\rho|i\>$.

Notice that only the first term of $\delta_{\mathcal{S},\mathcal{G}}$ depends on $\{ P_i \}$ and $\{ D_i \}$.
If $\rho_i\in\mathcal{V}$ for all states $\rho_i \in \mathcal{S}$,
reminding Eq. (\ref{eq-obsproj}) the first term of Eq. (\ref{eq-figmersg})
becomes
\begin{equation}\label{e:eta0}
\eta=\sum_i \Bra{D_i}Q_\mathcal{V} G Q_\mathcal{V}\Ket{D_i} \Tr[\rho_\mathcal{S} P_i].
\end{equation}

We now generalize this approach to tomography of quantum operations, keeping generally different
input and output Hilbert spaces $\hilb{H}_{in}$ and $\hilb{H}_{out}$, respectively. This has the
advantage that the usual tomography of states comes as the special case of one-dimensional
$\hilb{H}_{in}$, whereas tomography of POVMs corresponds to one-dimensional $\hilb{H}_{out}$.

A quantum operation is a trace non increasing CP-map $\mathcal{T}: \mathcal{B}(\hilb{H}_{in})
\longrightarrow \mathcal{B}(\hilb{H}_{out})$.  In order to gather information about a quantum
operation $\mathcal{T}$, the most general procedure consists in: \emph{i}) preparing a state $\rho \in
\mathcal{B}(\hilb{H}_{in}\otimes\hilb{H}_A)$ where $\hilb{H}_A$ is an ancillary
system with the same dimension of $\hilb{H}_{in}$; \emph{ii}) measuring the state $( \mathcal{T} \otimes
{\mathcal I}_A)(\rho)$ with a POVM $\{ P_i \}$. The probability of obtaining a generic outcome $i$ is given by
\begin{equation}
p_i = \Tr[(  \mathcal{T} \otimes\mathcal{I}_A) (\rho) P_i ],
\end{equation}
which, using the Choi-Jamio\l kowski isomorphism \cite{choi}, 
\begin{equation}
 \mathcal{T}(\rho)=\Tr_{in}[(I_{out} \otimes \rho^T)R_\mathcal{T}],
\;\,
R_\mathcal{T} = \mathcal{T} \otimes I_{in} (\KetBra{I}{I})
\end{equation}
becomes
\begin{equation}
\Tr[\Tr_{in}[(I_{A} \otimes R_{{\mathcal T}})
(\rho^{\theta_{in}} \otimes I_{out})]P_i]
=\Tr[R_{{\mathcal T}} \Pi_i^{(\rho)}],
\label{eqtester2}
\end{equation}
where  $\theta$ is
the transposition w.r.t. the orthonormal basis in Eq. (\ref{eq-corr}),
and
\begin{equation}\label{eqtester}
\Pi_i^{(\rho)} =\{\Tr_A[(\rho \otimes I_{out})(I_{in}
\otimes P_i^{\theta_{out}}) ]\}^T.
\end{equation}
It is convenient to use here the notion of \emph{tester} along with
the theoretical framework introduced in \cite{tester-papers}. A
tester is the natural generalization of the concept of POVM from
states to transformations, and is represented by a set of positive
operators $\{ \Pi_i \}$ with 
\begin{equation}
\sum_i \Pi_i = I \otimes \sigma, \qquad
\Tr[\sigma] = 1
\end{equation}

The probability distribution in Eq. (\ref{eqtester2}) is precisely represented by a Born-rule with
the tester $\{ \Pi_i \}$ in place of $\{ P_i \}$, and the operator $R_{{\mathcal T}}$ in place of
$\rho$. Such generalized Born rule can be rewritten in terms of the usual one as follows
\cite{tester-papers}
\begin{equation}
p_i=\Tr[R_{{\mathcal T}}\Pi_i]=\Tr[{\mathcal T}\otimes{\mathcal I}(\nu)P_i],
\end{equation}
with 
\begin{equation}\label{eq-testreal}
  \nu = \KetBra{\sqrt{\sigma}}{\sqrt{\sigma}}, \qquad P_i = (I\otimes \sigma^{-1/2})\Pi_i (I\otimes \sigma^{-1/2}).
\end{equation}
This method allows a straightforward generalization of the tomographic
method from states to transformation. Now tomographing a quantum
operation means using a suitable tester $\Pi_i$ such that the
expectation value of any other possible measurement can be inferred by
the probability distribution $p_i = \Tr[R_{{\mathcal T}} \Pi_i]$.  In
order to achieve this task we have to require that $\{ \Pi_i \}$ is
an operator frame for 
 $\mathcal{B}(\hilb{H}_{out} \otimes \hilb{H}_{in})$.
This means that  we can expand any operator on $\hilb{H}_{out} \otimes \hilb{H}_{in}$
 as follows
\begin{equation}\label{eq-operator}
  A = \sum_i \BraKet{\Delta_i}{A} \Pi_i \qquad A \in \mathcal{B}(\hilb{H}_{out} \otimes  \hilb{H}_{in}).
\end{equation}
where $\{ \Delta_i \}$ is a possible dual of the frame $\{ \Pi_i \}$,
that is the condition $\sum_i \KetBra{\Pi_i}{\Delta_i} = I_{out}
\otimes I_{in}$ holds.

Optimizing the tomography of quantum operations means minimizing the
statistical error in the determination of the expectation of a
generic operator $A$ as in Eq. (\ref{eq-operator}). This is provided
by the variance
\begin{equation}
 \delta(A) = \sum_i |\BraKet{\Delta_i}{A}|^2 \Tr[R_{{\mathcal T}} \Pi_i] - |\Tr[R_{{\mathcal T}} A]|^2
\end{equation}

We assume an ensemble $\mathcal{E} = \{R_k, p_k \}$ of possible
transformations and a weighted set $\mathcal{G} = \{A_n, q_n \}$ of
possible observables.  Averaging the statistical error over these
ensembles we obtain
\begin{equation}
\delta_{\mathcal{E},\mathcal{A}} := \sum_i \Bra{\Delta_i}G\Ket{\Delta_i}
 \Tr[R_\mathcal{E} \Pi_i] - \sum_{k,n} p_k q_n |\Tr[R_k A_n]|^2.
\end{equation}
Optimizing this figure of merit means:
  \emph{i}) optimizing the choice of the dual frame $ \{ \Delta_i  \}$;
  \emph{ii}) optimizing the choice of the frame $ \{ \Pi_i  \}$.
The optimization of the set $ \{ \Pi_i  \}$
reflects in  both choosing the best input state for the quantum operation
and the best final measurement.

In the following, for the sake of clarity we will consider
$\hbox{dim}(\hilb{H}_{in})=\hbox{dim}(\hilb{H}_{out})=:d$, and focus on the ``symmetric'' case $G = I$;
this happens for example when the set $\{ A_n \}$ is an orthonormal basis, whose elements are
equally weighted. Moreover, we assume that the averaged channel of the ensemble $\mathcal{E}$ is the
maximally depolarizing channel, whose Choi operator is $R_\mathcal{E}=d^{-1}I\otimes I$.

With these assumptions the relevant term of figure of merit becomes
\begin{equation}\label{e:eta}
  \eta = \sum_i \BraKet{\Delta_i}{\Delta_i}d^{-1}\Tr[\Pi_i].
\end{equation}

Since $R_\mathcal{E}$ is invariant under the action of $SU(d) \times
SU(d)$ we now show that it is possible to impose the same covariance
also on the tester without increasing the value of $\eta$.  Let us
define
\begin{align} \label{eq-testcovar}
\Pi_{i,g,h}& := (U_g\otimes V_h) \Pi_i (U_g^\dag \otimes V_h^\dag),\\
\Delta_{i,g,h}& := (U_g\otimes V_h) \Delta_i (U_g^\dag \otimes V_h^\dag).
\end{align}
It is easy to check that $\Delta_{i,g,h}$ is a dual of $\Pi_{i,g,h}$
by evaluating the group average after the sum on $i$. Then we observe
that the normalization of $\Pi_{i,g,h}$ gives
\begin{equation}
\sum_i \! \int \!\! dg dh \; \Pi_{i,g,h} = d^{-1}I \otimes I
\end{equation}
corresponding to $\sigma= d^{-1}I$ in Eq. (\ref{eq-testreal}), namely
one can choose $\nu=d^{-1}\KetBra{I}{I}$. In the last identity $dg$ and $dh$ are invariant measures
normalized to unit.

It is easy to verify that the figure of merit for the covariant tester
is the same as for the non covariant one, whence, w.l.o.g. we optimize
the covariant tester. The condition that the covariant tester is
informationally complete w.r.t. the subspace of transformations to be
tomographed will be verified after the optimization.

We note that a generic covariant tester is obtained by Eq.
(\ref{eq-testcovar}), with operators $\Pi_i$ becoming ``seeds'' of the
covariant POVM, and now being required to satisfy only the
normalization condition 
\begin{equation} \label{eq-norm}
\sum_i \Tr[\Pi_i]= d 
\end{equation}
 (analogous of covariant POVM normalization in
\cite{infocompgroup,holevo}).
The problem of optimization of the dual frame 
has been solved in \cite{ODP}. With the optimal dual, the figure of
merit simplifies as
\begin{equation}
\eta = \Tr[\tilde{X}^{-1}],
\end{equation}
where
\begin{align}
 \tilde{X}  = \sum_i \int \!\! dgdh \; \frac{ d \KetBra{\Pi_{i,g,h}}{\Pi_{i,g,h}} }{\Tr[\Pi_{i,g,h}]}=
\int \!\! dgdh \; W_{g,h}
 X W_{g,h}^\dag
\end{align}
with $W_{g,h}= U_g \otimes U^*_g \otimes V_h \otimes V^*_h$ and $X = \sum_i d\KetBra{\Pi_i}{\Pi_i}/\Tr[\Pi_i]$.
Using Schur's lemma we have \cite{note1}
\begin{align}
&\tilde{X}=P_1 + A P_2 + B P_3 + C P_4,\label{eq-projcov}\\
&\!\!\begin{array}{ll}
P_1= \Omega_{13} \otimes \Omega_{24},
&P_2= \left(I_{13} - \Omega_{13}\right) \otimes \Omega_{24},\\
P_3= \Omega_{13} \otimes \left(I_{24} - \Omega_{24}\right),
&P_4= (I_{13} - \Omega_{13}) \otimes (I_{24} - \Omega_{24}),\nonumber
\end{array}
\end{align}
having posed $\Omega= {\KetBra{I}{I}}/{d}$ and
\begin{align}
&A = \frac{1}{d^2-1} \left\{\sum_i\frac{\Tr[(\Tr_2[\Pi_i])^2]}{\Tr[\Pi_i]}-1\right\}\nonumber\\
&B = \frac{1}{d^2-1} \left\{\sum_i\frac{\Tr[(\Tr_1[\Pi_i])^2]}{\Tr[\Pi_i]}-1\right\}\\
&C = \frac{1}{(d^2-1)^2}\left\{\sum_i \frac{d\Tr[\Pi_i^2]}{\Tr[\Pi_i]}-(d^2-1)(A+B)-1\right\}.\nonumber
\end{align}
One has
\begin{equation}\label{eq-trx}
\Tr[\tilde{X}^{-1}]= 1 + (d^2-1) \left( \frac{1}{A}+\frac{1}{B} + \frac{(d^2-1)}{C}\right).
\end{equation}

We note that if the ensemble of transformations is contained in a
subspace $\mathcal{V} \subseteq
\mathcal{B}(\hilb{H}_{out}\otimes\hilb{H}_{in})$ the figure of merit
becomes $\eta= \Tr[\tilde{X}^\ddagger Q_\mathcal{V}]$,
where $\tilde{X}^\ddagger$ is the Moore-Penrose pseudoinverse. We now
carry on the minimization for three relevant subspaces:
\begin{align}
& \mathcal{Q} = \mathcal{B}(\hilb{H}_{out}\otimes\hilb{H}_{in}),\qquad
\mathcal{C} = \{ R \in \mathcal{Q}, \; \Tr_{out}[R]=I_{in} \} & \nonumber  \\
& \mathcal{U} = \{ R\in \mathcal{Q} , \; \Tr_{out}[R]=I_{in}, \Tr_{in}[R]=I_{out} \}&
\end{align}
corresponding respectively to quantum operations, general channels and
unital channels. The subspaces $\mathcal{C}$ and $\mathcal{U}$ are
invariant under the action of the group $\{ W_{g,h} \}$ and thus the
respective projectors decompose as
\begin{equation}
Q_\mathcal{C} = P_1 + P_2 + P_4, \qquad Q_\mathcal{U} = P_1 + P_4
\end{equation}

\begin{figure}[htbp]
\begin{center}
\includegraphics[width=240pt]{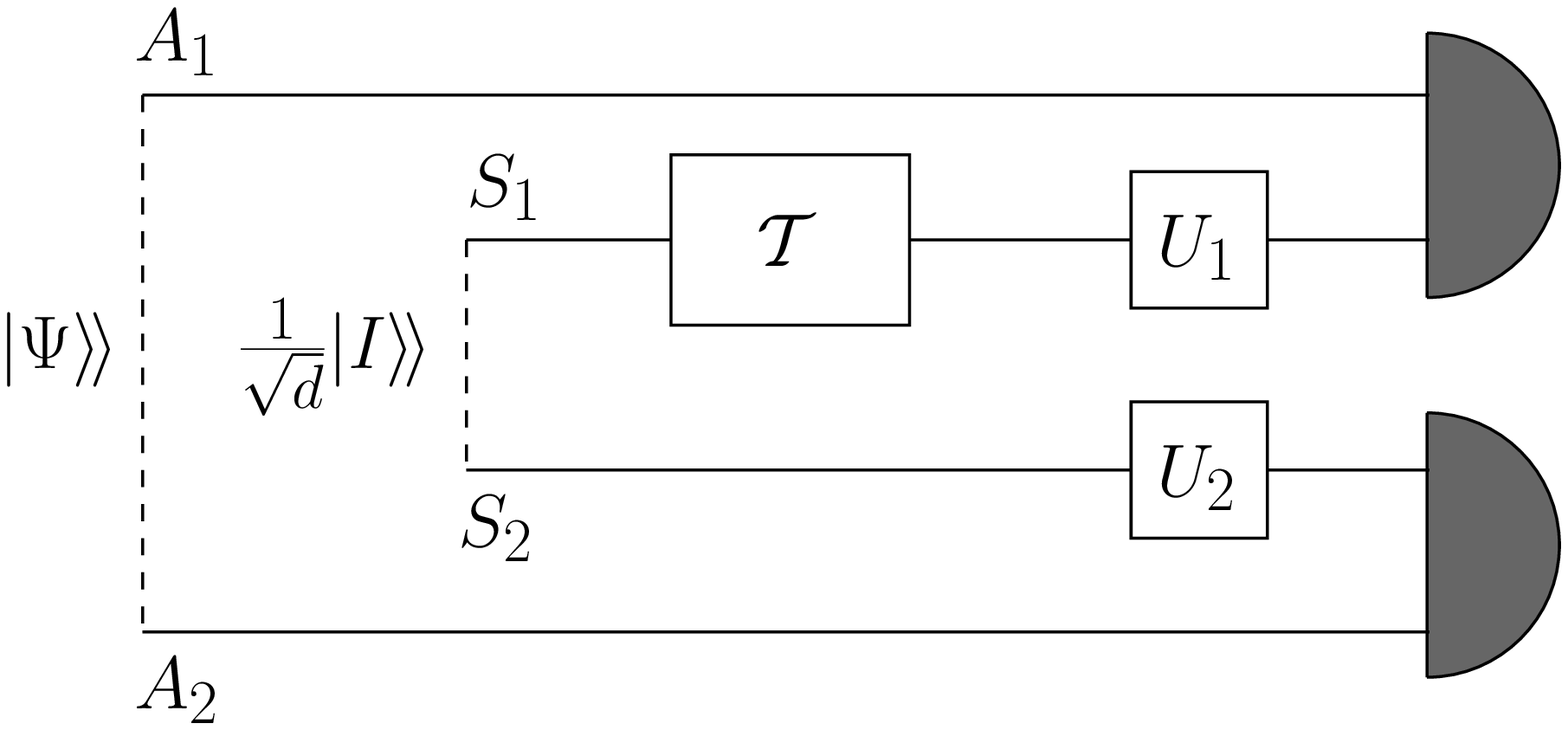}
\caption{Physical implementation of optimal quantum transformation
  tomography.  The two measurements are Bell's measurements preceded by a random unitary. The state $\Ket{\Psi}$ depends on the prior ensemble.
\label{fig}
}
\end{center}
\end{figure}

Without loss of generality we can assume the operators $\{ \Pi_i \}$
to be rank one. In fact, suppose that $\Pi_i$ has rank higher than
1. Then it is possible to decompose it as $\Pi=\sum_{j}\Pi_{i,j}$ with
$\Pi_{i,j}$ rank 1. The statistics of $\Pi_i$ can be completely
achieved by $\Pi_{i,j}$ through a suitable post-processing. For the
purpose of optimization it is then not restrictive to consider rank
one $\Pi_i$, namely $\Pi_i=\alpha_i\Ket{\Psi_i}\Bra{\Psi_i}$, with
$\sum_i\alpha_i=d$.  Notice that all multiple seeds of this form lead
to testers satisfying Eq. (\ref{eq-norm}).

In the three cases under examination, the figure of merit is then
\begin{align}
&\eta_\mathcal{Q}=\Tr[\tilde X^{-1}]=1+(d^2-1)\left(\frac2A+\frac{(d^2-1)^2}{1-2A}\right) & \nonumber \\
&\eta_\mathcal{C}=\Tr[\tilde X^\ddagger Q_\mathcal{C}] =1+(d^2-1)\left(\frac1A+\frac{(d^2-1)^2}{1-2A}\right) & \nonumber \\
&\eta_\mathcal{U}=\Tr[\tilde X^\ddagger Q_\mathcal{U}] =1+(d^2-1)\left(\frac{(d^2-1)^2}{1-2A}\right) &
\end{align}
where $0\leq
A=(d^2-1)^{-1}(\sum_i\alpha_i\Tr[(\Psi_i\Psi_i^\dag)^2]-1) \leq\frac1{d+1}<\frac12$.
The minimum can simply be determined by derivation with respect to
$A$, obtaining $A=1/(d^2+1)$ for quantum operations,
$A=1/(\sqrt2(d^2-1)+2)$ for general channels and $A=0$ for unital
channels. The corresponding minimum for the figure of merit is
\begin{align}
&\eta_\mathcal{Q} \geq d^6+d^4-d^2 & \nonumber \\
&\eta_\mathcal{C} \geq d^6+(2\sqrt2-3)d^4+(5-4\sqrt2)d^2+2(\sqrt2-1) & \nonumber \\
&\eta_\mathcal{U} \geq (d^2-1)^3+1.&
\end{align}
The same result for quantum operations and for unital channels has been obtained in \cite{scott2} in a different framework.

These bounds are simply achieved by a single seed $\Pi_0=d\Ket{\Psi}\Bra{\Psi}$, with
\begin{equation}
 \Tr[(\Psi\Psi^\dag)^2] = \frac{2d}{d^2+1},\quad
 \frac{\sqrt2(d^2-1)+3}{d(\sqrt2(d^2-1)+2)},\quad 1 
\end{equation}
respectively for quantum operations, general channels and unital channels,
namely with
\begin{equation}
\Psi=[d^{-1}(1-\beta) I+\beta\ket{\psi}\bra{\psi}]^{\frac12}
\end{equation}
where
$\beta=\sqrt{(d+1)/(d^2+1)}$ for quantum operations, $\beta=[(d-1)(2+\sqrt2(d^2-1))]^{-1/2}$ for general channels and $\beta=0$ for unital channels,
and $\ket{\psi}$ is any pure state.
The informationally completeness is thus verified \emph{a posteriori}
 (see \cite{infocompgroup}).

The same procedure can be carried on when the operator $G$ has the more general form $G=g_1 P_1 + g_2 P_2 + g_3 P_3 + g_4 P_4$, where $P_i$ are the projectors defined in (\ref{eq-projcov}). In this case Eq. (\ref{eq-trx}) becomes
\begin{equation}
\Tr[\tilde{X}^{-1}G]= g_1 + (d^2-1) \left( \frac{g_2}{A}+\frac{g_3}{B} + \frac{(d^2-1)g_4}{C}\right),
\end{equation}
which can be minimized along the same lines previously followed.
$G$ has this form when optimizing measuring procedures of this kind:
\emph{i}) preparing an input state  randomly drawn from  the set $\{U_g \rho U^{\dag}_g\} $;
\emph{ii}) measuring an observable chosen from the set  $\{U_h A U^{\dag}_h\} $.

We now show how the optimal measurement can be experimentally implemented. Referring to Fig.
\ref{fig}, the bipartite system carrying the Choi operator of the transformation is indicated with
the labels $S_1$ and $S_2$. We prepare a pair of ancillary systems $A_1$ and $A_2$ in the joint
state $\KetBra{\Psi}{\Psi}$, then we apply two random unitary transformations $U_1$ and $U_2$ to
$S_1$ and $S_2$, finally we perform a Bell measurement on the pair $A_1 S_1$ and another Bell
measurement on the pair $A_2 S_2$. This experimental scheme realizes the continuous measurement by
randomizing among a continuous set of discrete POVM; this is a particular application of a general
result proved in \cite{contmeasure}. The scheme proposed is feasible using e.~g. the Bell
measurements experimentally realized in \cite{zeilinger}.  We note that choosing $\Ket{\Psi}$
maximally entangled (as proposed for example in \cite{mohseni}) is generally not optimal, except for
the unital case.

With the same derivation starting from Eq. (\ref{e:eta}), but keeping
$\mbox{dim}(\hilb{H}_{in})\neq\mbox{dim}(\hilb{H}_{out})$, one obtains the optimal tomography for
general quantum operations. The special case of $\text{dim}(\hilb{H}_{in})=1$ (one has $P_3=P_4=0$
in Eq. (\ref{eq-projcov})) corresponds to optimal tomography of states, whereas case
$\text{dim}(\hilb{H}_{out})=1$ ($P_2=P_4=0$) gives the optimal tomography of POVMs. The
corresponding experimental schemes are obtained by removing the upper/lower branch for
POVMs/states, respectively. In the remaining branch the bipartite detector becomes a mono-partite,
performing a von Neumann measurement for the qudit, preceded by a random unitary in $SU(d)$.
Moreover, for the case of POVM, the state $\Ket{\Psi}$ is missing, whereas, for state-tomography,
both bipartite states are missing. The optimal $\eta$ in Eq. (\ref{e:eta0}) is given by
$\eta=d^3+d^2-d$, in both cases (for state-tomography compare with Ref. \cite{scott1}).

In conclusion, we presented a general method for optimizing quantum tomography, based on the new
notion of {\em tester}. The method is very versatile, allowing to consider arbitrary prior ensemble
and representation. We provided the optimal experimental schemes for tomography of states and
various kinds of process tomography, giving the corresponding performance, all schemes being
feasible with the current technology.


\begin{thebibliography}{99}
\bibitem{note1}
We label the four spaces in such a way that $W_{g,h}\in \mathcal{B}(\hilb{H}_1\otimes\hilb{H}_3\otimes\hilb{H}_2\otimes\hilb{H}_4)$, with $\hilb{H}_1=\hilb{H}_{out}$ and $\hilb{H}_2=\hilb{H}_{in}$.
\end{thebibliography}
\end{document}